\documentclass[12pt,preprint]{aastex}

\shorttitle{Infrared Spectroscopy of NGC 891 Halo}
\shortauthors{Rand et al.}

\begin{document}

%% LaTeX will automatically break titles if they run longer than
%% one line. However, you may use \\ to force a line break if
%% you desire.

\title{Infrared Spectroscopy of the Diffuse Ionized Halo of NGC 891}

%% Use \author, \affil, and the \and command to format
%% author and affiliation information.
%% Note that \email has replaced the old \authoremail command
%% from AASTeX v4.0. You can use \email to mark an email address
%% anywhere in the paper, not just in the front matter.
%% As in the title, use \\ to force line breaks.

\author{Richard J. Rand}
\affil{Department of Physics and Astronomy, University of New
Mexico, 800 Yale Blvd, NE, Albuquerque, NM 87131}
\email{rjr@phys.unm.edu}
\and

\author{Kenneth Wood}
\affil{School of Physics and Astronomy, University of St. Andrews,
North Haugh, St. Andrews KY16 9SS, UK}
\email{kw25@st-andrews.ac.uk}

\author{Robert. A. Benjamin}
%\author{Robert. A. Benjamin\altaffilmark{1}}
\affil{Department of Physics, University of Wisconsin at Whitewater,
800 West Main Street, Whitewater, WI 53190}
\email{ benjamir@uww.edu}

%% Mark off your abstract in the ``abstract'' environment. In the manuscript
%% style, abstract will output a Received/Accepted line after the
%% title and affiliation information. No date will appear since the author
%% does not have this information. The dates will be filled in by the
%% editorial office after submission.

\begin{abstract}
We present infrared spectroscopy from the {\it Spitzer Space Telescope} at one
disk position and two positions at a height of 1 kpc from the disk in the
edge-on spiral NGC 891, with the primary goal of studying halo ionization.
Our main result is that the [Ne$\,$III]/[Ne$\,$II] ratio, which provides a
measure of the hardness of the ionizing spectrum free from the major problems
plaguing optical line ratios, is enhanced in the extraplanar pointings
relative to the disk pointing.  Using a 2D Monte Carlo-based photo-ionization
code which accounts for the effects of radiation field hardening, we find that
this trend cannot be reproduced by any plausible photo-ionization model, and
that a secondary source of ionization must therefore operate in gaseous halos.
We also present the first spectroscopic detections of extraplanar PAH features
in an external normal galaxy.  If they are in an exponential layer, very rough
emission scale-heights of $330-530$ pc are implied for the various features.
Extinction may be non-negligible in the midplane and reduce these
scale-heights significantly.  There is little significant variation in the
relative emission from the various features between disk and extraplanar
environment.  Only the 17.4 $\mu$m feature is significantly enhanced in the
extraplanar gas compared to the other features, possibly indicating a
preference for larger PAHs in the halo.

\end{abstract}

%% Keywords should appear after the \end{abstract} command. The uncommented
%% example has been keyed in ApJ style. See the instructions to authors
%% for the journal to which you are submitting your paper to determine
%% what keyword punctuation is appropriate.

%% Authors who wish to have the most important objects in their paper
%% linked in the electronic edition to a data center may do so in the
%% subject header.  Objects should be in the appropriate "individual"
%% headers (e.g., quasars: individual, stars: individual, etc.) with the
%% additional provision that the total number of headers, including each
%% individual object, not exceed six.  The \objectname{} macro, and its
%% alias \object{}, is used to mark each object.  The macro takes the object
%% name as its primary argument.  This name will appear in the paper
%% and serve as the link's anchor in the electronic edition if the name
%% is recognized by the data centers.  The macro also takes an optional
%% argument in parentheses in cases where the data center identification
%% differs from what is to be printed in the paper.


%% From the front matter, we move on to the body of the paper.
%% In the first two sections, notice the use of the natbib \citep
%% and \citet commands to identify citations.  The citations are
%% tied to the reference list via symbolic KEYs. The KEY corresponds
%% to the KEY in the \bibitem in the reference list below. We have
%% chosen the first three characters of the first author's name plus
%% the last two numeral of the year of publication as our KEY for
%% each reference.

\section{Introduction}

One of the most important changes in our way of thinking about interstellar
gas in spiral galaxies in the past twenty years or so derives from the
discovery that the ISM is much thicker than had been previously thought.
Layers of great vertical extent (referred to as ``extraplanar gas'' or
``gaseous halos'') have been found in just about every component of the ISM,
especially in X-rays (e.g., \citealt{2006A&A...448...43T,2004ApJS..151..193S}),
HI (e.g., \citealt*{2007AJ....134.1019O}), radio continuum
(e.g., \citealt*{2006A&A...457..121D}), diffuse ionized gas (DIG;
e.g., \citealt{2003A&A...406..505R,1996ApJ...462..712R}), and dust
(\citealt{1999AJ....117.2077H,1998ApJ...507L.125A,2000A&AS..145...83A}).
This gas may originate in star-formation driven disk-halo flows
(e.g., \citealt{1989ApJ...345..372N}), primordial infall
(\citealt*{1997ApJ...477..765C}) or accretion from satellite galaxies
(\citealt{2005ASPC..331..139V}).  Evidence for a connection to star formation
comes from the correlation of the brightness and extent of extraplanar DIG,
X-ray, radio continuum and dust with the level of underlying disk star
formation, both within and among galaxies
(\citealt{1996ApJ...462..712R,1998PASA...15..106R,
2003A&A...406..493R,2006A&A...457..779T,2004ApJS..151..193S,2006A&A...457..121D,
1999AJ....117.2077H}).  On the other hand, some Galactic High Velocity Clouds
have metallicities as low as 0.1 $Z_{\sun}$ \citep{2004Ap&SS.289..381W}, which is
not expected for gas originating in the disk.

\subsection{Emission Lines from DIG}

In the Milky Way, the DIG (or Reynolds Layer) accounts for most of the ionized
ISM \citep{1990ApJ...349L..17R}, and represents a significant power
requirement.  Energetically, only photo-ionization by massive stars can
comfortably explain the intensity of optical line emission
\citep{1990ApJ...349L..17R}, which raises the question of how an ionizing
source concentrated in a thin disk can be responsible for such thick ionized
layers \citep{1993ApJ...417..579M,1994ApJ...430..222D}.

Much information regarding the source(s) of ionization have come from optical
emission line ratios in the Milky Way and external galaxies (e.g., \citealt*
{1997ApJ...474..129R,1998ApJ...501..137R,1997ApJ...483..666G,2001ApJ...551...57C,
2002ApJ...572..823O,1999ApJ...523..223H,2003ApJ...586..902H,2006ApJ...652..401M}).
Studies of the spatial behavior of several diagnostic line ratios have
supported the idea that photo-ionization by massive stars is the primary
ionization mechanism, but crucial aspects of the data require a more complex
picture.  The most characteristic result is that the ratios of
[S$\,$II]$\lambda\lambda 6716,6731$ and [N$\,$II]$\lambda\lambda 6548,6583$ to
H$\alpha$ generally increase with distance from the suspected ionizing source,
in the Milky Way and in external galaxy disks and halos.  This was initially
interpreted as a fall in the ionization parameter, $U$, with distance from the
ionizing source (e.g., \citealt{1994ApJ...428..647D,1998ApJ...501..137R}) but
more recently has been attributed to an increase in the gas temperature in the
DIG relative to the immediate star forming environment due to a source of
extra non-ionizing heating
(e.g., \citealt{1999ApJ...523..223H,2002ApJ...572..823O}, using further
information from [O$\,$II$]\lambda 3727$/H$\alpha$ in edge-ons), while even
more recently the importance of radiation field hardening during propagation
(due to the longer mean free path for harder photons) has been stressed
(\citealt{2003ApJ...586..902H,2004MNRAS.353.1126W}, hereafter WM).

Measurements of other line ratios present further difficulties for a simple
photo-ionization picture.  In several external spiral galaxies, both edge-on
and more face-on, [O$\,$III]$\lambda 5007$/H$\beta$ is found to increase, or
at least vary little, with distance from the ionizing massive stars (e.g.,
\citealt*{1998ApJ...501..137R, 2000ApJ...537L..13R,1999AJ....118.2775G};
\citealt{2002ApJ...572..823O, 2003ApJ...586..902H,2003ApJ...592...79M}).  An
increase has also been found in the Reynolds Layer for gas near the tangent
point near $l=27^{\circ}$ (\citealt{2005ApJ...630..925M}).  This behavior is
unexpected for dilute radiation fields and has led to the consideration of
secondary sources of ionization, such as shocks, in many of the
above-referenced studies.  Extra non-ionizing heating in halos may also be a
factor, although \citet{2001ApJ...551...57C} find that this is an unlikely
explanation in NGC 5775 and UGC 10288.  Exceptions to this behavior of
[O$\,$III]/H$\beta$ do exist, however
\citep{2000A&A...362..119T,2003ApJ...592...79M}.  [O$\,$I]$\lambda
6300$/H$\alpha$, which depends on the neutral fraction of H through a charge
exchange reaction, when detected in the halos of edge-ons is generally
enhanced relative to the disk
\citep{1998ApJ...501..137R,2000ApJ...537L..13R,2003ApJ...592...79M}.  It is
very weak in the Reynolds Layer \citep{1998ApJ...494L..99R}, and has only been
detected in one external non-edge-on spiral, M33 \citep{2006ApJ...644L..29V}.
He$\,$I$\lambda 5876$ is difficult to detect, but He$\,$I/H$\alpha$ traces the
relative ionization fractions of helium and hydrogen and thus is a useful
measure of the hardness of the radiation field.  In the Milky Way,
\citet{2006ApJ...652..401M} have detected the line in a few diverse
environments, finding a range of helium ionization fractions, but in general
the diffuse gas is in a lower ionization state than in HII regions.  The line
is detected in M33 with large uncertainties
\citep{2003ApJ...586..902H,2006ApJ...644L..29V}, while in the halo of NGC 891
helium is about 70\% ionized \citep{1997ApJ...474..129R}.

Spectroscopy of the DIG in irregular galaxies has led to a similar picture,
with massive stars being the main source of ionization, but shocks possibly
contributing as well [based on observations of 15 such galaxies
by \citet{1997ApJ...475...65H} and \citet{1997ApJ...491..561M}].

In summary, then, although radiation from massive stars dominates the
ionization of DIG, many other factors may be at play: non-ionizing heating,
shocks or other secondary ionization sources, and radiation field hardening
during propagation.  In addition, photo-ionization structure will depend on
the how the ionization parameter changes with distance from the massive stars,
and on the mean stellar spectrum emerging from the star forming regions.
Further complications arise from uncertainties in the halo gas abundances, and
also of course from dust extinction, which is not insignificant even at
distances 1--2 kpc from the midplane of several edge-on spirals, including
particularly NGC 891 \citep[][and references
therein]{1999AJ....117.2077H,2000AJ....119..644H}.

The Infrared Spectrograph \citep[IRS;][]{2004ApJS..154...18H} on board the {\it
Spitzer Space Telescope} \citep{2004ApJS..154....1W} opens a unique window on
the ionization of diffuse gas through its access to infrared gas-phase lines,
in particular the 12.81 $\mu$m [Ne$\,$II] and 15.55 $\mu$m [Ne$\,$III] lines,
and in this paper we present mid-infrared spectra of the disk and halo of NGC
891.  The ratio of these lines provides a diagnostic of ionization that is
relatively insensitive to extinction, gas-phase abundances, and temperature
(being low excitation lines in warm gas) -- three of the biggest sources of
confusion for the optical lines.  The first two ionization potentials (IPs) of
Ne are 21.6 and 41.0 eV.  With fewer complications, this ratio should provide
excellent constraints on whether the diffuse gas is purely photo-ionized or
requires an additional source by comparing against models.  The utility of
these lines as ionization diagnostics in Galactic and extragalactic HII
regions and starbursts was demonstrated using the {\it Infrared Space
Observatory} (ISO) by, e.g., \citet{2000ApJ...539..641T},
\citet{2002ApJ...566..880G}, and \citet{2003A&A...403..829V}.

The aforementioned modeling by WM uses the 3D Monte-Carlo radiative transfer
code of \citet*{2004MNRAS.348.1337W} to address several of the issues
associated with photo-ionization.  This code is able to employ more realistic
geometries for the ambient gas -- such as clumpy or fractal ISM distributions
that may have a great impact on the ionization structure -- than can be done
by the 1D codes previously applied to the DIG problem (e.g.,
\citealt{1994ApJ...428..647D}).  It also incorporates radiation field
hardening, non-ionizing heating, and a range of possible input spectra and
luminosities.  WM find that spectral hardening reduces the need for the
non-ionizing heating source invoked by, e.g., \citet{1999ApJ...523..223H}, by
producing higher gas temperatures and ratios of forbidden lines to Balmer
lines, while suppressing He$\,$I emission.  However, it is still not possible
to match all of the line ratio data for the Reynolds layer and the halo of NGC
891 without invoking additional heating or a secondary source of ionization.
In this paper, we apply models tailored to the diffuse ISM density
distribution of NGC 891 to constrain the sources of ionization using existing
optical line ratios and the unique information provided by our {\it Spitzer}
data.

\subsection{Emission from PAH Features}

Although our primary focus is on gas-phase lines, {\it Spitzer} spectroscopy
also provides access to the mid-infrared spectral features that are generally
attributed to polycyclic aromatic hydrocarbons (PAHs;
\citealt*{1984A&A...137L...5L,1985ApJ...290L..25A,2001ApJ...556..501B};
\citealt{2001ApJ...560..261B}; \citealt*{2007ApJ...657..810D}).  The
identification of individual large molecules, or families of such molecules,
responsible for the various features, as well as their creation, destruction,
heating and emission in different environments are all topics of current
interest.  \citet*{2004ApJ...613..986P} argue that PAH emission in general
should be a good tracer of B stars.  While PAH emission should be broadly
connected to star formation, it is suppressed relative to the emission from
larger grains in the immediate vicinity of recently formed stars
\citep[e.g,][]{2007ApJ...665..390L}.  This is usually attributed to
destruction of PAH molecules by harsh UV radiation fields.
\citet{2006ApJ...639..157W} find that the equivalent widths (EW) of several
PAH features are lower in BCDs than in typical starburst galaxies, and find a
weak anticorrelation of PAH EWs with [Ne$\,$III]/[Ne$\,$II].  A stronger
anticorrelation is found with a parameter given by this ratio multiplied by
the UV luminosity and divided by the metallicity.  Thus, hardness, radiation
field intensity and metallicity may all bear on PAH chemistry and emission.
In galaxy halos the hardness, at least in the 20--40 eV range, is traced by
the [Ne$\,$III]/[Ne$\,$II] ratio.  The starlight intensity will clearly be
weaker than in disks.  As PAH emission is dominated by single-photon heating,
their spectrum is not expected to vary with starlight intensity for the
kinds of environments relevant here \citep{2007ApJ...657..810D}.

Further modification of the dust grain population may occur through
sputtering, grain collisions, and other processes \citep{2003ARA&A..41..241D}.
For instance, \citet*{1996ApJ...469..740J} find that shocks, with speeds as
low as 50 km s$^{-1}$, could be a source of PAHs through shattering of larger
grains.  Radiative acceleration of grains, with a size-dependent efficiency,
may also cause a modification of the grain size population
\citep{1991ApJ...381..137F}, with smaller grains staying closer to their
formation sites \citep{1998MNRAS.300.1006D}.  These effects could lead to
differences in the grain population between disks and halos [see also
\citet{1997AJ....114.2463H} for a general discussion of processes that may
lead to halo dust].  The vertical extent of PAH emission in halos and possible
changes in the relative strengths of PAH features with height, and how these
vary in galaxies with different star formation rates, could shed light on the
chemistry of these large molecules.

Here we report spectroscopic detections of PAHs at 11.2, 12.0, 12.7, 16.5, and
17.4 $\mu$m, and a broad plateau from 16.5 to 17.5 $\mu$m in the halo of an
external normal spiral, for the first time (the first spectroscopic PAH
detections in any external halo were by \citet{2006ApJ...642L.127E} for the
starburst galaxy M82).  PAHs emitting in this wavelength range are thought to
be generally neutral and larger than PAHs emitting shortward of 10 $\mu$m
\citep{2007ApJ...657..810D}.  Features at 11--14 $\mu$m are thought to arise
from C-H out-of-plane bending modes.  The strong 11.2 $\mu$m feature is
believed to arise from compact, condensed PAHs, while the 12.7$\mu$m feature
is thought to arise from more irregularly shaped molecules
\citep{2004ARA&A..42..119V}.  The red tail of the 11.2 $\mu$m feature is
thought to arise mainly from cations \citep{1999ApJ...516L..41H}.
\citet{2004ApJ...611..928V} find that the shape of the 11.2 $\mu$m feature, in
terms of the asymmetry of the blue and red sides, and the strength of the
smaller feature at 11.0 $\mu$m, varies with the type of exciting source (i.e.,
HII regions, Herbig AeBe stars, young stellar objects and planetary nebulae).

Features in the 15--20 $\mu$m region have been discovered only recently.
\citet{2000A&A...354L..17M} first reported the 16.5 $\mu$m feature, in
Galactic sources, while the 17.4 $\mu$m feature can be seen in the spectra of
some of the Galactic sources observed by \citet{2000A&A...357.1013V} and was
also tentatively seen in the nuclei of M82 and NGC 253 by
\citet{2000A&A...358..481S}.  A broad plateau from 15 to 20 $\mu$m was found
by \citet{2000A&A...357.1013V}.  \citet{2004ApJ...613..986P} explore how these
features change with the type of exciting source (i.e., HII regions, young
stellar objects and evolved stars), and conclude that the discrete features in
this wavelength range arise generally from prominent bands in neutral PAHs
that should be larger than those responsible for the 10--15 $\mu$m features.
while plateaus are due to blending of less prominent bands.  They also find
that the 17.4 $\mu$m feature may arise from several candidate PAHs.  In the
dust model of \citet{2007ApJ...657..810D}, the PAHs responsible for
this feature are also very large (300--10$^4$ C atoms).  Emission in
this wavelength range from C-C-C bending modes of larger PAHs was predicted by
\citet*{1989ApJS...71..733A}.

As far as previous observations of PAH emission from galaxy halos are
concerned, \citet{2006A&A...445..123I}, using archival ISO data for the
edge-on NGC 5907, detect mid-infrared narrow-band emission well into the halo.
They argue that PAH features should dominate the emission in the filters used,
and thus that the PAHs in this halo have a typical scale-height of 3.5 kpc,
with emission extending up to 6.5 kpc.  NGC 5907 is a galaxy with a low star
formation rate and little evidence for a DIG halo \citep{1996ApJ...462..712R}.
This result is therefore rather unexpected and needs to be confirmed.

Finally, PAH features from 6.2 to 11.2 $\mu$m in the disk of NGC 891 have
already been detected by \citet*{1999A&A...342..643M}, who find variation in
the band ratios along the major axis, suggesting changes in the
dehydrogenation degree and charge of the PAHs.

\section{Observations}

The data were taken on 2006 February 2 and 2006 March 4 (program ID
20380; PI: R. Rand) using
%(\dataset{ADS/Sa.Spitzer\#14595584}\dataset{ADS/Sa.Spitzer\#14595840},
%\dataset{ADS/Sa.Spitzer\#14596096}\dataset{ADS/Sa.Spitzer\#14596352},
%\dataset{ADS/Sa.Spitzer\#14596608}\dataset{ADS/Sa.Spitzer\#14596864},
%\dataset{ADS/Sa.Spitzer\#14597120}\dataset{ADS/Sa.Spitzer\#14597376})
the staring mode of the IRS Short-High (SH) module on board {\it Spitzer}.  A
log of the observations is given in Table 1.  The SH module is a
cross-dispersed echelle spectrograph providing spectral coverage from 9.9
to 19.6 $\mu$m with a resolving power of $\approx$ 600.  The aperture
has dimensions of $4.7\arcsec$x$11.3\arcsec$.  In staring mode the SH module
nods between two pointings centered 1/3 and 2/3 of the way along the slit.

% Actual slit centers at 970 and 1120 pc on E and W.  Full extent of nodded
% slits covers 700 pc.

Deep observations were made at locations in the halo of NGC 891 centered at a
height of approximately $z= \pm 1$ kpc (hereafter ``east halo'' and ``west
halo'' pointings) above a location in the disk at $R=100\arcsec$ from the
center of the galaxy.  The reconstructed pointings indicate that the actual
slit centers were at heights of 970 and 1120 pc on the east and west sides,
respectively.  A shorter observation of the disk location was also obtained.
The exact coordinates of the pointings are given in Table 1, and the pointings
are shown in Figure 1.  The disk pointing falls between HII regions visible in
Figure 1, although more embedded HII regions may contribute to the emission.
The long integrations for the halo pointings meant that the data were divided
into three Astronomical Observation Requests (AORs) per pointing, while the
disk observation required only one AOR.  Because of the bright infrared
background, separate sky observations close in time to the target observations
are necessary for IRS exposures of faint emission, particularly when the
emission is expected to fill the slit.  A deep sky exposure was taken on 2006
March 4 as part of this program.  Unfortunately, the integration time is not
as long as for the halo observations, resulting in the sky exposure dominating
the noise in the halo spectra.  No suitable sky exposure exists for the 2006
February 2 observations.  However, only a small minority of the data were
taken on this date (see Table 1) and since all lines of interest are detected
with good sensitivity, these data were not used in the analysis.

\clearpage
\begin{figure}
\epsscale{.80}
\includegraphics[scale=1,viewport=50 0 574 468]{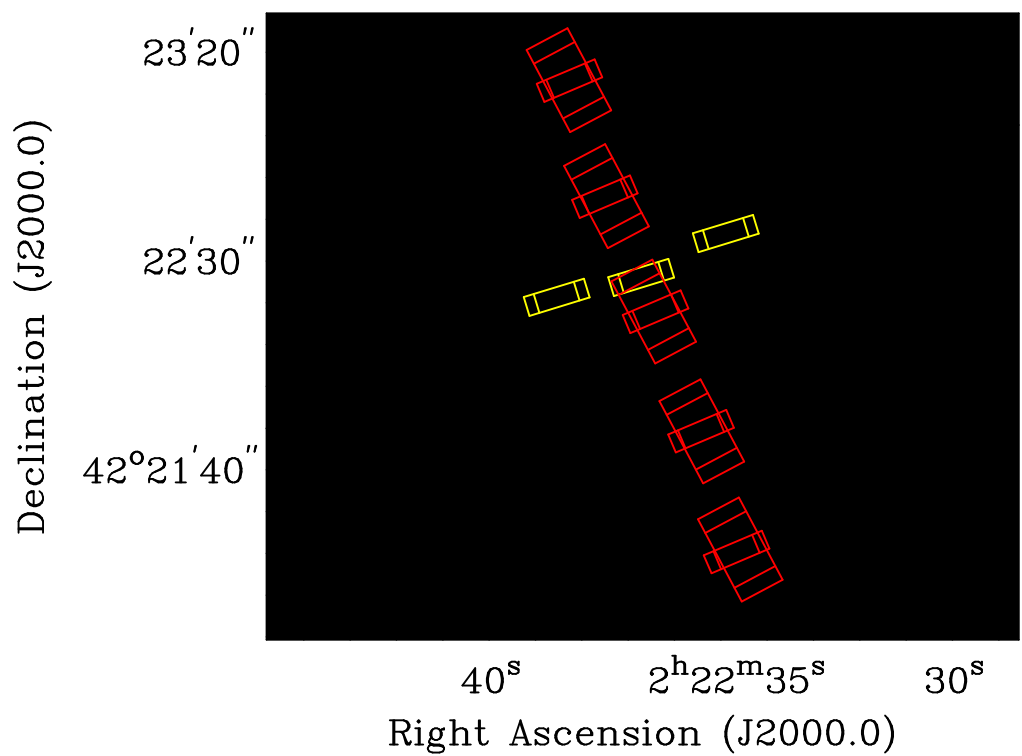}
%Black and white version for print version
%\includegraphics[scale=1,viewport=50 0 574 468]{f1bw.eps}
\caption{Section of an H$\alpha$ image of NGC 891 \citep{1990ApJ...352L...1R}.
  Boxes show IRS pointings (for each pointing, two overlapping boxes, showing
  the two nods, are drawn).  The three (yellow) boxes closest to the long-slit
  are the IRS SH pointings discussed here.  The other (red) boxes in the
  disk show pointings for the SH and LH GTO data discussed.  The solid line
  shows the orientation of the slit for the optical emission line data
  discussed.  The slit width is $2.25\arcsec$.  The white blotch in the
  H$\alpha$ image is an artifact of continuum subtraction of a foreground
  star.
\label{fig1}}
\end{figure}
\clearpage

The data were processed through version S13.2 of the IRS reduction pipeline,
providing the Basic Calibrated Data (BCD) products.  Post-BCD processing was
carried out according to the Infrared Spectrograph Data Handbook Version 2.0.
Charge accumulation can be present in long integrations, evidenced by steady
increases in the signal, but no significant effect was found in our data.  The
sky observation was subtracted from all target observations, and ``rogue''
pixels cleaned using the contributed IRSCLEAN\_MASK software.  For each of the
AORs listed in Table 1 and for each nod, the 2-D spectra were averaged using
sigma-clipping at the 3$\sigma$ level.  All remaining extreme pixels not
removed by IRSCLEAN\_MASK were replaced with averages of adjacent pixels.
One-dimensional spectra were extracted using the package SPICE (version 1.3).
Emission from NGC 891 is apparent over the entire aperture in every pointing,
so the spectra were extracted and calibrated (with calibration tables, based
on standard star observations, included in SPICE) using the full aperture
width.  Note that while version S13.2 of the pipeline allows extended source
extraction, it is only fully accurate if the source uniformly fills the
slit. The exact corrections for other emission distributions are impossible to
estimate unless the source structure is known in advance.  However, our
comparison of our disk spectrum with nearby GTO pointings (see below)
demonstrate that any such corrections will not affect the main results of this
paper.  A final spectrum for each pointing was formed by combining the
resulting 1-D spectra for each AOR and nod.  As the spectral orders overlap,
the ends of each order were removed from the final spectra.  The spectra are
shown in Figure 2.

\clearpage
\begin{figure}
\epsscale{.80}
\includegraphics[scale=.8,viewport=0 100 574 768]{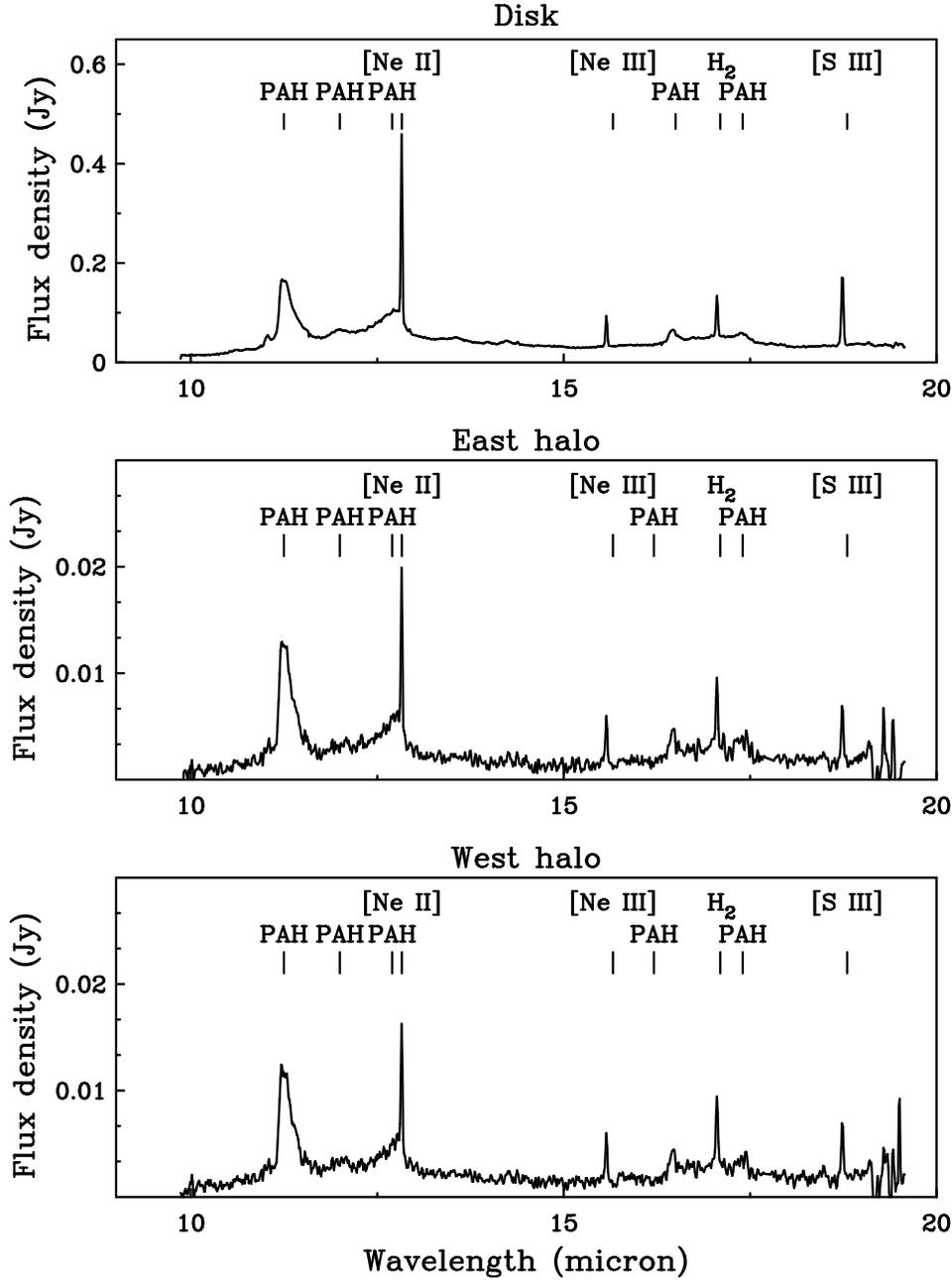}
\caption{IRS SH spectra of the disk and east and west halo fields of NGC 891.
Detected gas and dust phase features are indicated.
\label{fig2}}
\end{figure}
\clearpage

As we only had one disk pointing, we checked whether our gas-phase emission
line ratios were representative by examining several pointings near our disk
position in the Spitzer archive from GTO data, taken on
2004 August 7 (program ID 97; PI: J. Houck).
%(\dataset{ADS/Sa.Spitzer\#4935936}) 
Both SH and LH (covering 18.7--37.2 $\mu$m at a resolving power of $\approx
600$ with an aperture of $11.1\arcsec$ x$22.3\arcsec$) data were reduced.
These pointings, spaced by about $11\arcsec$ in R.A. and 28$\arcsec$ in Decl.,
are also shown in Figure 1.  Integration times per nod are 31.5 and 14.7 s for
the SH and LH modules, respectively.  We refer to these pointings as A to E,
with A being southernmost.  No dedicated sky observations exist for these
pointings, so we used sky observations from 2004 August 8 (program ID 148).
%(\dataset{ADS/Sa.Spitzer\#9777152})
Because of the differing ecliptic latitude of these observations, the spectra
had to be scaled to the expected sky level for the NGC 891 pointings using the
predicted sky levels provided by the SPOT software before subtraction.
Otherwise, the data were processed as above.

Line intensities and EWs were measured with the IRAF\footnote[1]{IRAF is distributed
by the National Optical Astronomy Observatory, which is operated by the
Association of Universities for Research in Astronomy, Inc., under cooperative
agreement with the National Science Foundation.} program {\it splot} by
summing pixels over the line extent and subtracting a linear baseline.  Error
bars on intensities are based on the noise in the spectra and the estimated
uncertainties in the flux scale of IRS data, as described in the Infrared
Spectrograph Data Handbook Version 2.0.  Of the various sources of uncertainty
in the flux scale, most affect the overall scaling independent of wavelength,
while one is dependent on the position of the line in the order, and can be as
much as 5\%.  We assume conservatively that it is everywhere at its maximum
value.  The total systematic error on intensities is then about 9\%.  The
random noise in a single pixel is of order 10$^{-17}$ erg cm$^{-2}$ s$^{-1}$.
For error bars on line ratios and EWs, we use only the random noise in the
spectra and the wavelength-dependent error.

We checked that the fluxes in our halo pointings cannot be due to disk
emission convolved with the PSF of the IRS SH module using the {\it Spitzer}
contributed software STINYTIM.  The emission in the disk pointing
is peaked around the midplane.  At a representative wavelength of 17 $\mu$m,
we estimate that a point source in the midplane would contribute 0.04\% of its
peak intensity in our halo apertures due to the wings of the PSF, whereas the
halo emission we detect for all features is at least 3\% of the disk emission.
Therefore we expect very little contamination from the disk in the halo
pointings.

We also show optical emission line ratios in Figure 3 from the long-slit
spectra of \citet{1998ApJ...501..137R}.  The observations are described
in that paper.

\section{Results}

\subsection{Ionized Gas Phase Emission Lines}

In all three pointings, we detect [Ne$\,$II] 12.81 $\mu$m, [Ne$\,$III] 15.56
$\mu$m, and [S$\,$III] 18.71 $\mu$m (Figure 2; Table 2).  [S$\,$III] has an IP
of 23.3 eV, similar to [Ne$\,$II].  We have also placed upper limits on
[S$\,$IV] 10.51 $\mu$m emission.  All of our disk line intensities fall
between those for GTO pointings C and D (Table 3), as would naively be
expected from the positions.  To the extent that these limited measurements
can be used to find the equivalent scale-height of an assumed exponential
distribution of emission, the range of such scale-heights would be about
$280-350$ pc for [Ne$\,$II] and [S$\,$III], and $380-480$ pc for [Ne$\,$III],
given the error bars, and assuming no extinction (but see below).  These
change little if the average of the five intensities in the GTO data are used
instead as a representative disk intensity.  The [Ne$\,$III]/[Ne$\,$II] ratio
is $0.13 \pm 0.01$ in the disk and $0.31 \pm 0.03$ and $0.32 \pm 0.03$ in the
east and west halo, respectively (see also Figure 3a).  For the five GTO
pointings A--E, the [Ne$\,$III]/[Ne$\,$II] ratios are 0.12, 0.13, 0.14, 0.13
and $<$0.13, respectively.  For [S$\,$III]/[Ne$\,$III], we find ratios of $2.1
\pm 0.2$ in the disk and $0.8 \pm 0.1$ and $0.9 \pm 0.1$ in the east and west
halo respectively.  For GTO pointings A--E, the ratios are 1.9, 1.9, 2.0, 2.8
and $<$ 3.2, respectively.  Hence, our disk results seem representative of the
general trend in that part of the disk, while the two sides of the halo give
similar ratios to each other.  There is therefore a significant variation of
these line ratios between the disk and the halo.  Such variations we will
refer to as halo-disk contrasts.  There is no halo-disk contrast in
[S$\,$III]/[Ne$\,$II], within the errors, consistent with their nearly equal
IPs.  The [Ne$\,$III]/[Ne$\,$II] results indicate a higher ionization state in
the halo than in the disk.  In \S{4} we explore whether this contrast can be
explained by pure photo-ionization or whether a second source of ionization is
required.  We simply note here that the result cannot be explained by an
additional non-ionizing source of heating in a pure photo-ionization model, as
this ratio is temperature independent.

In a perfectly edge-on, gas rich disk such as NGC 891, extinction in the
midplane may be significant even at mid-IR wavelengths.  To estimate this
effect, we calculate total gas column densities from the CO map of
\citet{1993ApJ...404L..59S} and the HI map of \citet{2007AJ....134.1019O}.  We
calculate H$_2$ column densities from CO intensities assuming a conversion
factor of $3 \times 10^{20}$ mol cm$^{-2}$ (K km s$^{-1}$)$^{-1}$, as used by
\citet{1993ApJ...404L..59S}.  We assume the Galactic relation between $A_V$
and gas column density from \citet{1978ApJ...224..132B}, convert $A_V$ to
$A_K$ using a standard Galactic reddening law
\citep[e.g.,][]{2003ARA&A..41..241D}, and finally employ a value of $A_{15
\mu{\rm m}}/A_K$ of 0.4, consistent with determinations for many directions in
the Galactic plane by \citet{2006A&A...446..551J}.  Extinction in this part of
the spectrum is very uncertain and depends on the strength of the 18 $\mu$m
silicate absorption feature, which, while clear in the ISO spectrum of the
Galactic Center from \citet{1999ESASP.427..623L}, is not evident in our disk
spectrum.  Nevertheless, with these inputs we calculate about 2 magnitudes of
extinction at 15 $\mu$m at our disk position.  Extinction at a $z=1$ kpc
is negligible.  The possible effect of reddening on
[Ne$\,$III]/[Ne$\,$II] is even less well constrained.  If we use the
extinction at 12 $\mu$m towards the Galactic Center from
\citet{1999ESASP.427..623L}, which exceeds that at 15 $\mu$m, the disk neon
ratio would be further lowered.

Both disk and halo [Ne$\,$III]/[Ne$\,$II] ratios are typical of nearby
starburst galaxies studied by \citet{2000ApJ...539..641T}, and much lower than
the HII regions 30 Doradus and W51 these authors compare against -- a contrast
they attribute to aging effects.  However, these authors find that DIG is not
likely to contribute much to their global ratios for starbursts.  With
increasing distance to the starburst, the $14\arcsec$x$27\arcsec$ ISO SWS
aperture should encompass a greater fraction of diffuse gas.  This led to the
expectation of a decrease in the line ratio with starburst distance under the
assumption of a lower ratio for DIG, and such a decrease is not observed by
them.  However, if the ratio generally increases with distance from ionizing
stars, as found here, but with still relatively low values in the DIG, then
such a trend might not be expected.

\clearpage
\begin{figure}
\epsscale{.80}
\includegraphics[scale=.8,viewport=0 100 574 768]{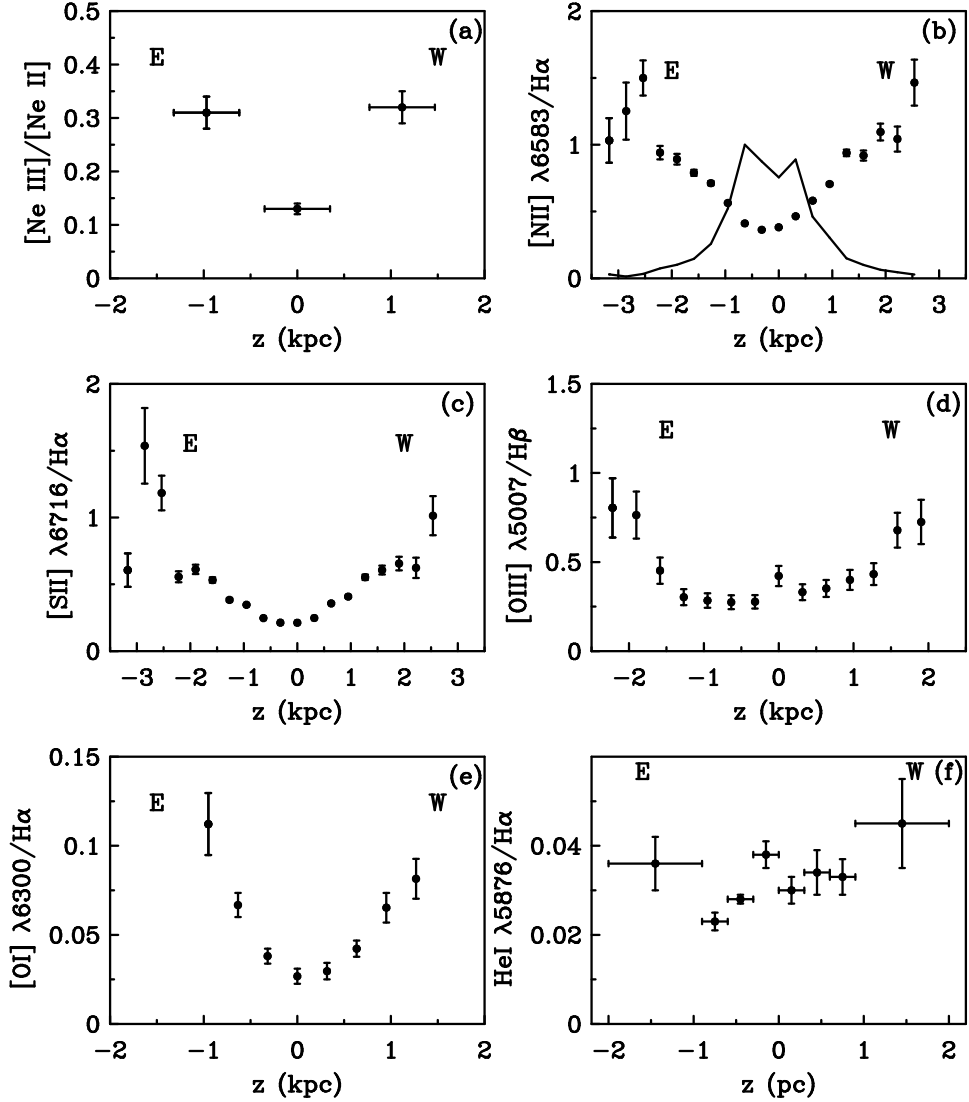}
\caption{Dependence of (a) [Ne$\,$III]/[Ne$\,$II], (b) [N$\,$II]$\lambda
6583$/H$\alpha$, (c) [S$\,$II]$\lambda 6716$/H$\alpha$, (d) [O$\,$I]$\lambda
6300$/H$\alpha$ (e) [O$\,$III]$\lambda 5007$/H$\beta$, and (f) He$\,$I
$\lambda 5876$/H$\alpha$ on $z$.  The optical line
ratios are from \citet{1997ApJ...474..129R} and \citet{1998ApJ...501..137R}.
The H$\alpha$ profile, normalized to unit intensity, from
\citet{1998ApJ...501..137R}, is shown in (b).  Horizontal error bars in (a)
and (f) reflect the extent over which the data have been averaged.  In the
other panels, the intensities were averaged over 317 pc.
\label{fig3}}
\end{figure}
\clearpage

\subsection{PAH Emission Features}

We detect five PAH features, at 11.2, 12.0, 12.7, 16.5 and 17.4 $\mu$m, and a
broad plateau from roughly 16.5 to 17.5 $\mu$m (Figure 2; Table 2).  The
halo-disk emission contrasts of these five features as ordered above are
$0.068 \pm 0.006$, $0.04 \pm 0.02$, $0.048 \pm 0.011$, $0.056 \pm 0.003$, and
$0.14 \pm 0.03$.  Most features therefore have similar contrasts within the
errors, consistent with the expectation that the PAH spectrum should not vary
much with starlight intensity as long as it is not very strong.  Only the 17.4
$\mu$m feature stands out as having a relatively high halo-disk contrast.  The
contrast for this line still stands out if the upper limits for the GTO data
are used for the disk intensities (although upper limits on some other
features complicate the comparison).  It is also the only feature with a
significantly higher EW in the halo (0.12 and 0.079 vs.  0.038), suggesting a
larger vertical extent for this feature relative to the small, warm grains
responsible for the continuum.  Averaging the east and west halo measurements,
the equivalent scale-heights of an assumed exponential distribution of
emission would be about $330-530$ pc for these features.  For two magnitudes
of extinction in the disk, the range becomes $210-270$ pc.  The scale-heights
change little if we use instead the average of the disk intensities in the GTO
observations (Table 3).  For the 11.2 and 12.7 $\mu$m features (the only ones
detected in all five GTO pointings), they become 460 and 380 pc, respectively,
assuming no extinction.  We also note for comparison that
\citet{1999A&A...342..643M} found an intensity of the 11.2 $\mu$m feature of
$2000 \pm 350$ erg cm$^{-2}$ s$^{-1}$ arcsec$^{-2}$ for a
$24\arcsec$x$24\arcsec$ field centered at our disk position using ISO,
somewhat larger than our value of $1400 \pm 100$ erg cm$^{-2}$ s$^{-1}$
arcsec$^{-2}$, although the aperture sizes are not matched (again, the
IRS SH aperture area is 53.11 arcsec$^2$).

CO emission at this location has a narrower vertical distribution.  It is
found to have a FWHM of about 225 pc in the high-resolution observations of
\citet{1993ApJ...404L..59S}, which would correspond to an exponential
scale-height of 160 pc.  Modeling of the HI disk by
\citet{2007AJ....134.1019O} reveals a thin disk with a scale-height of $<$0.3
kpc in the inner disk, rising to 0.5 kpc in the outer regions, and a thick
disk with 30\% of the HI mass, with a scale-height increasing with radius from
1.25 kpc to 2.5 kpc.  The PAH features may therefore have a vertical
distribution comparable to the main HI layer, but the extinction correction
would make their distribution intermediate between the CO and HI layers.

\citet{2006A&A...445..123I} modeled the vertical distribution of emission in
NGC 5907 in the ISO 6.7W filter with one to three components.  It is
impossible to say whether there is one or indeed more vertical components to
the PAH distribution from our data, but nevertheless one can state that if the
PAH distribution in NGC 891 is represented by a single exponential
distribution, then the range of FWHMs for the features of 450--730 pc,
assuming no extinction, is less than that of even the narrowest component
(830 pc) in NGC 5907 (in a model with three Gaussian layers).  With our data
we of course cannot rule out an extended tail to the PAH emission distribution
with a scale-height of a few kpc as found for NGC 5907 by Irwin \& Madden, but
such an extended halo seems very unlikely in the more actively star-forming
NGC 891.

Because four of the five PAH features show no significant halo-disk EW
contrast, there is obviously no correlation with [Ne$\,$III]/[Ne$\,$II].  Note
that we are only spanning a small range, and a lower regime, of this ratio
compared to the BCD sample of \citet{2006ApJ...639..157W}, where a weak
correlation was found.  However, the 17.4 $\mu$m EW is {\it higher} in the
halo, where [Ne$\,$III]/[Ne$\,$II] is higher -- the opposite of the weak trend
found by Wu et al. for the 6.2 $\mu$m and 11.2 $\mu$m features.  The lack of
EW halo-disk contrasts also puts limits on the modification of the grain size
distribution by processes such as sputtering, shattering and radiative
acceleration as discussed in \S 1.

Also as discussed in \S 1, the origin of the 17.4 $\mu$m feature is uncertain
-- it could arise from several relatively large, neutral, candidate PAHs
\citep{2004ARA&A..42..119V}.  The higher halo-disk contrast for this line may
therefore indicate a halo favoring larger PAHs.  Shattering of larger grains
by shocks \citep{1996ApJ...469..740J} in the halo is a potential source of
such molecules.

The spectral resolution of the IRS SH module permits an analysis of the shapes
of PAH emission feature profiles.  The profiles of the 11.2 $\mu$m feature,
after subtraction of a linear continuum interpolated across regions adjacent
to the line, are shown in Figure 4 (the two sides of the halo have been
averaged).  The shape of this feature is found to vary with environment in the
Milky Way by \citet[see their Figure 3]{2004ApJ...611..928V}.  Most common is
the Class A$_{11.2}$ profile (relatively steep rise of the blue edge and fall
of the red tail), typical of HII regions, non-isolated (i.e., associated with
reflection nebulosity) Herbig AeBe stars and young stellar objects), while the
Class B$_{11.2}$ (shallower rise of the blue edge, slightly redward shifted
peak, shallower red tail, and the 11.0 $\mu$m secondary feature shifted
redward and broadened) is shown by many planetary nebulae and post-AGB
stars. A few sources feature a hybrid class A(B)$_{11.2}$ profile (similar to
A$_{11.2}$ but with a shallower redward falloff and a slight redward shift of
the 11.0 $\mu$m feature).  In Figure 4, there is no discernible difference
between the disk and halo profiles, suggesting little variation in the
neutral/ionized mix or the excitation modes of the family of PAHs likely
responsible for this feature.  The profiles most resemble the A$_{11.2}$
type, with a relatively steep blue rise, shallow redward falloff, and a
clear secondary feature at 11.05 $\mu$m.  The location of the peak of the
secondary feature is more similar to that in the A(B)$_{11.2}$ profile.
However, the main point is that there is no evidence for significant evolution
of the PAHs responsible for the feature between the disk and halo.

\clearpage
\begin{figure}
\epsscale{.80}
\includegraphics[scale=.8,viewport=0 50 574 668]{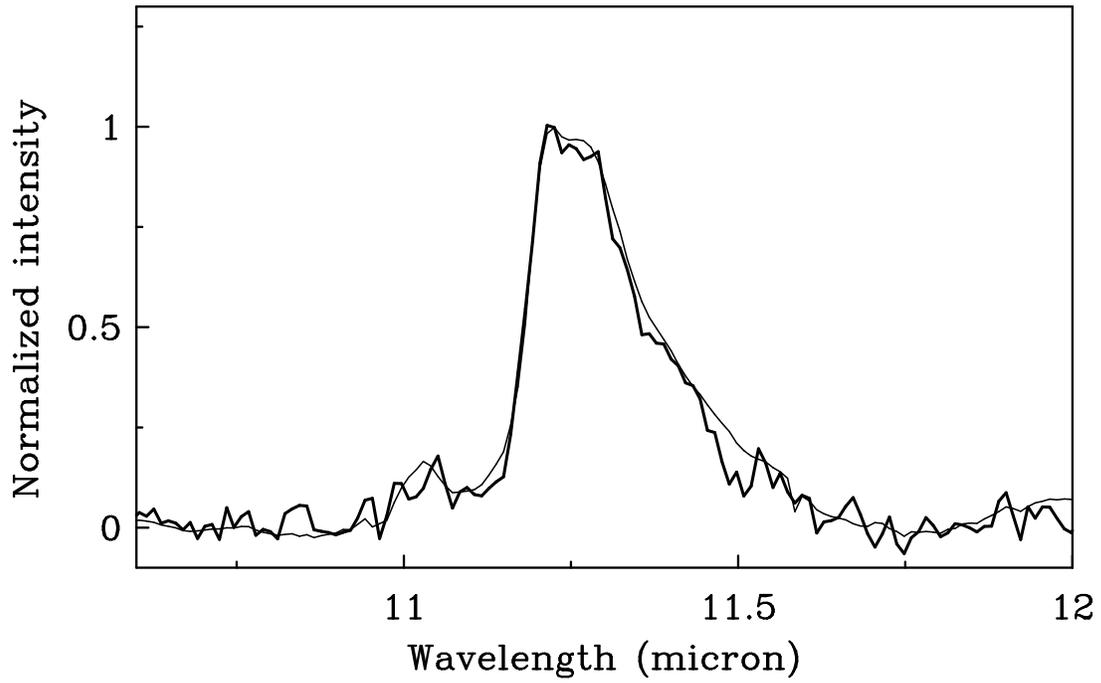}
\caption{The 11.2 $\mu$m feature for the disk (solid line) and the average of the east
and west halo pointings (dotted line).  A linear continuum has been subtracted
from the spectra.
\label{fig4}}
\end{figure}
\clearpage

The 15--18 $\mu$m emission features a broad plateau on which the 16.4 and 17.4
$\mu$m features sit.  This part of the spectrum, after subtraction of a linear
continuum using regions adjacent to the plateau, is shown in Figure 5 for the
disk and the average of the two halo pointings.  The location of the plateau,
and the strengths of the two features relative to it and to each other, is
much more reminiscent of the YSOs studied by \citet{2004ApJ...613..986P} than
the HII region in their sample.  There is also a hint in the disk spectrum of
the blue end of the second broad plateau seen in the spectrum of the ridge of
H$_2$ near the Herbig Be star LKH$\alpha$234 by Peeters et al. from 18 to 20
$\mu$m, here seen as a shallow rise from 18 $\mu$ to 18.7 $\mu$m.  Peeters et
al. can reproduce both the HII region- and YSO-type profiles with mixes of
laboratory PAH spectra, but the mixtures are not unique.

\clearpage
\begin{figure}
\epsscale{.80}
\includegraphics[scale=.8,viewport=0 50 574 668]{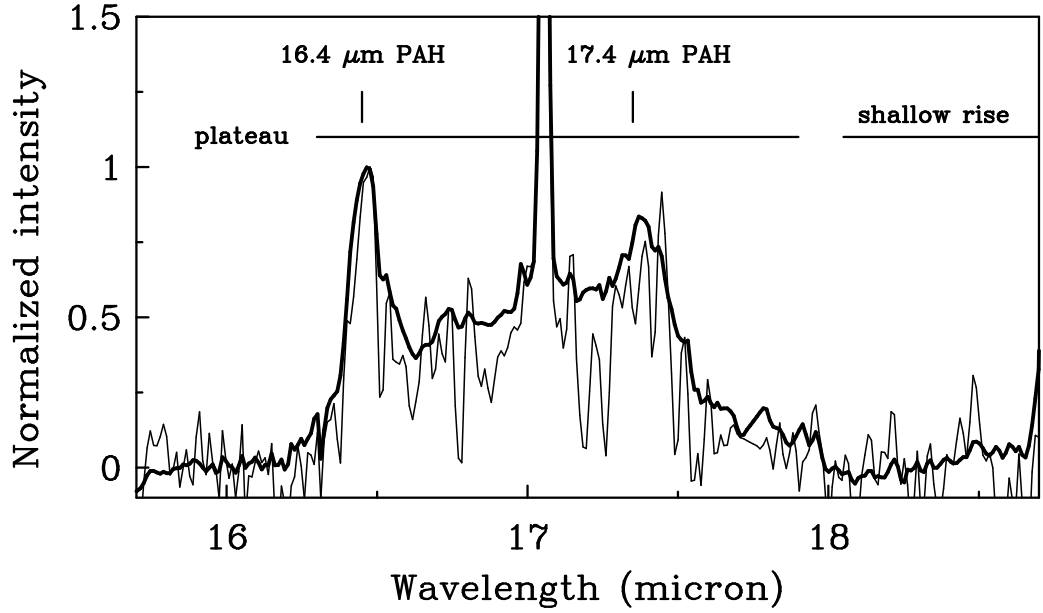}
\caption{The 15--20 $\mu$m region for the disk (thick line) and the average of the
east and west halo pointings (thin line).  A linear continuum has been subtracted
from the spectra.
\label{fig5}}
\end{figure}
\clearpage

\subsection{H$_{\rm 2}$ Emission}

H$_2$ mid-IR emission lines arise from rotational excitation in warm gas.  It
is usually attributed to PDRs \citep[e.g.,][]{2004ARA&A..42..119V} but can
also arise in shocks \citep[e.g.,][]{2006ApJ...649..816N}.  Such emission has
been mapped in the disk of NGC 891 using ISO Short-Wavelength Spectrometer
(SWS) by \citet{1999ApJ...522L..29V}.  We detect the H$_2$ {\it S}(1) $J=3-1$
17.03 $\mu$m rotationally excited line in the disk and halo.  Our disk
intensity is about 1.7 times higher than found for this line at the same
position in the galaxy by \citet{1999ApJ...522L..29V}, but the ISO SWS
aperture size is not well matched to the IRS SH slit, being about seven times
larger.  Following the above analysis for the PAH features, if the emitting
molecular hydrogen is in an exponential layer, then the measured halo-disk
contrast suggests a scale-height of $390-460$ pc.  This result changes little
if the GTO values are used.  Again, extinction may lower the scale-height
significantly: using our extinction estimate in \S 3.1 would bring the
scale-heights down to $220-260$ pc, still slightly thicker than the
CO-emitting layer.

\section{Ionization Modeling}

Figure 3 includes not only the [Ne$\,$III]/[Ne$\,$II] ratio but also the
optical line ratios along the long-slit (Figure 1) from
\citet{1998ApJ...501..137R}.  We wish to understand if these line ratios can
be reasonably reproduced by photo-ionization from a population of massive
stars residing in the disk.  We focus on models that predict at least
semi-quantitatively the observed neon ratio first, given the aforementioned
advantages of this ratio as an ionization diagnostic, and then examine how
well such models reproduce the optical line ratios.

We calculated photo-ionization models using the 3D Monte Carlo
photo-ionization code described in \citet{2004MNRAS.348.1337W}.  This code
calculates the ionization and temperature structure for 3D distributions of
gas and ionizing sources for elements H, He, C, N, O, Ne, and S.  We do not
consider ions with ionization potentials above 54~eV (the second ionization
potential of He), therefore we do not consider He$^{2+}$ or radiation transfer
of photons with energies above 54~eV.  This simplification is justified by the
lack of detection of He$^{2+}$ emission in the Reynolds layer and only its
rare detection in HII regions.  The models do not include the effects of dust
on either the transfer of ionizing photons or the resulting line ratios.
Observationally, from the CO and HI maps, modeling of the gas layers, and
assumptions about extinction in \S 3.1, $A_V \approx 3$ mag at 500 pc above a
position 100$\arcsec$ north of the galaxy center.  Below this height, then,
observed optical ratios may be somewhat affected by reddening.  For instance,
at 500 pc, the most reddened ratio, He$\,$I/H$\alpha$, may be reddened by
about 0.7 mag.  Other ratios are for lines much more closely separated in
wavelength.  For the bulk of the optical emission line data points in Figure
3, then, reddening should be insignificant and will not affect our
conclusions.  The simulations can include heating over and above that from
photo-ionization, according to any given prescription (see WM), but we have
not included additional heating in these simulations.  The elemental
abundances we adopt by number relative to H are: H/He = 0.1, C/H = 140 ppm,
N/H = 75 ppm, O/H = 319 ppm, Ne/H = 117 ppm, and S/H = 18.6 ppm.  With the
exception of S, these are the gas phase abundances in the local ISM used by
\citet{2000ApJ...544..347M}.  The S abundance we use was found by WM to better
match Milky Way and NGC 891 data.  Note that the calculated S line ratios are
highly uncertain due to the unknown dielectronic recombination rates.  In our
simulations we use the dielectronic recombination rates for S suggested by
\citet{1991PASP..103.1182A}.

All models have a central ionizing source and ionize a simulation volume that
extends to $\pm 1$~kpc in $x$, $y$, and $z$, with 65 cubical grid cells on a
side.  The ionizing spectra are 40kK, 45kK, and 50kK model atmospheres from
the WM-Basic library computed by \citet*{2003ApJ...599.1333S}.  The standard
ionizing luminosity is $10^{51}$ H-ionizing photons per second, except in one
model where we consider a luminosity four times higher.  Also, for one
simulation, we use a "leaky" model in which the input spectrum is that from a
spherical 40kK simulation which has its physical size set to allow 15\% of the
ionizing photons to escape.  This leaky model suppresses the He-ionizing
photons with energies above 24.6~eV and hardens the spectrum in the H-ionizing
continuum --- see Figure~11 of WM and the accompanying discussion there.

The density structure within the simulation grid is given by
\begin{equation}
n(z) = 0.15 \exp[-|z| / 0.66] + 0.04 \exp[-|z|/2.3]
\end{equation}
where the densities are in cm$^{-3}$ and distances are in kpc.  This
is a representative structure only, which we base on the HI data of
Oosterloo et al. (2007) and \citet{1997ApJ...491..140S}.  Our main
results will not depend strongly on this assumed distribution.

We note that the models predict [O$\,$III]/H$\alpha$ while our data are for
[O$\,$III]/H$\beta$.  However, extinction should not affect the observed
[O$\,$III]/H$\beta$ since the lines are well matched in wavelength.  Hence we
expect trends with $|z|$ for [O$\,$III]/H$\alpha$ in the models to be
comparable to those in [O$\,$III]/H$\beta$ in the data.

\clearpage
\begin{figure}
\epsscale{.80}
\includegraphics[scale=.8,viewport=-100 -50 574 568]{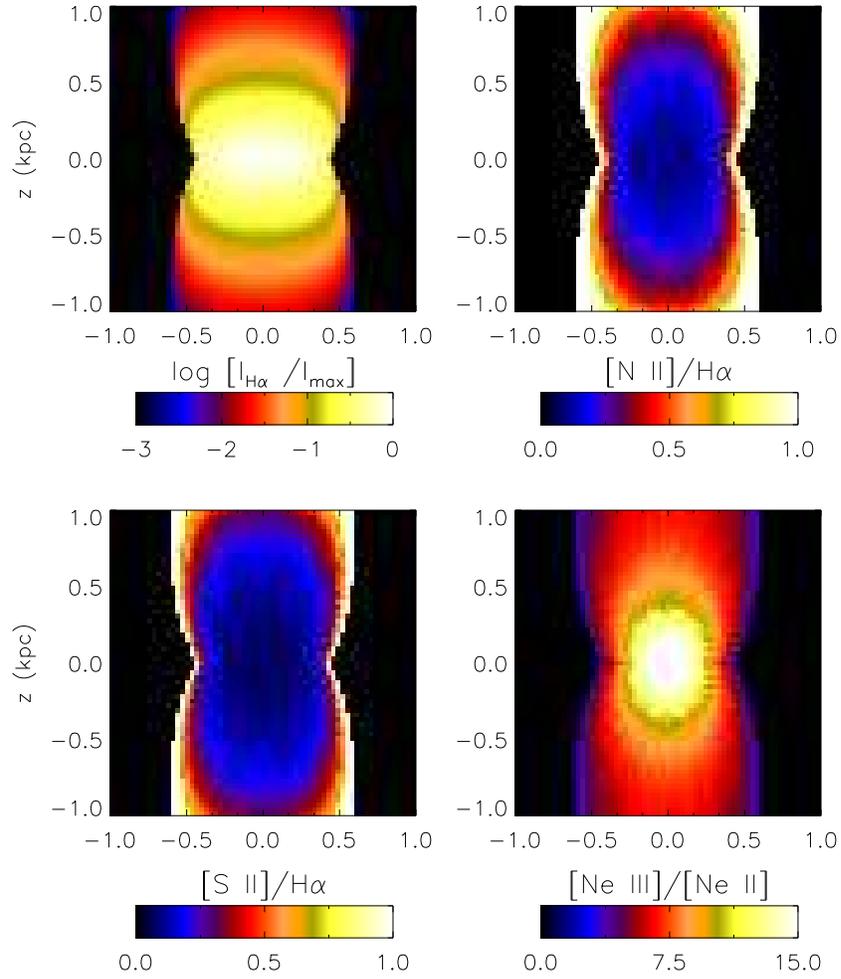}
%Black and white version for print version
%\includegraphics[scale=.8,viewport=-100 -50 574 568]{f6_bw.eps}
\caption{Total projected relative intensity and line ratio maps for H$\alpha$,
[N$\,$II]/H$\alpha$, [S$\,$II]/H$\alpha$ and [Ne$\,$III]/[Ne$\,$II] for a
model with a 40kK ionizing source of luminosity $10^{51}$ H-ionizing photons
per second.
\label{fig6}}
\end{figure}

\begin{figure}
\epsscale{.80}
\includegraphics[scale=0.6]{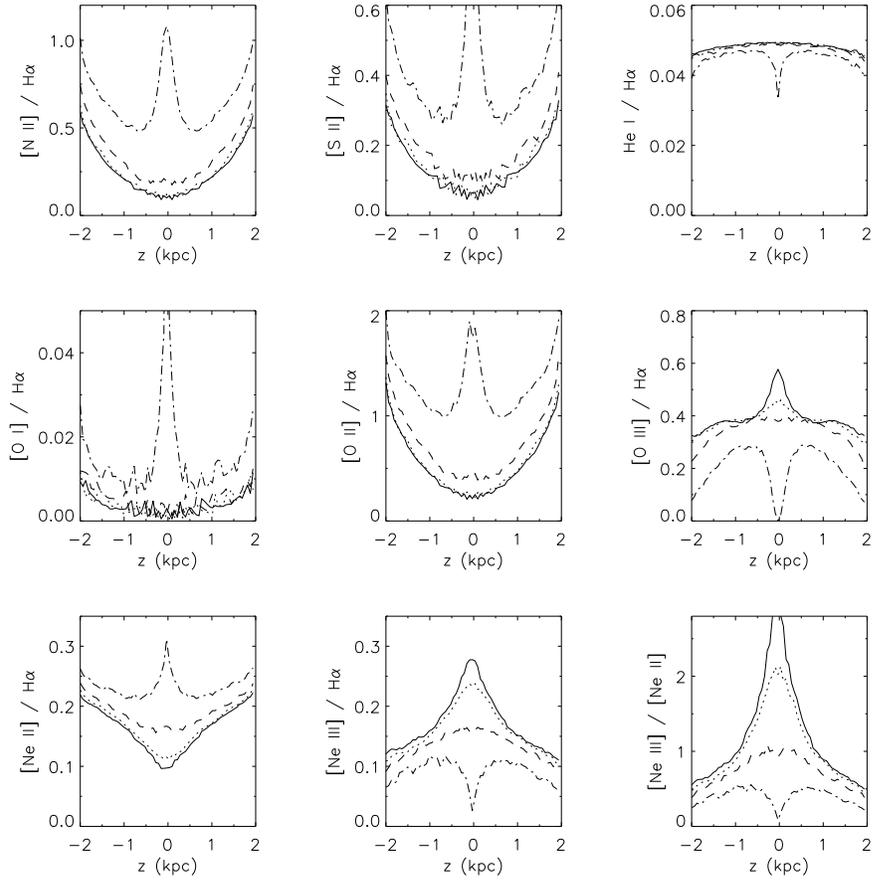}
\caption{Vertical cuts at $x=$ 0 kpc (solid), 0.3 kpc (dotted), 0.6 kpc (dashed)
and 0.9 kpc (dot-dashed) showing the variation of line ratios with $z$ for
the model in Figure 6.
\label{fig7}}
\end{figure}
\clearpage

We illustrate trends in the models by first showing in Figure 6 2D maps of the
H$\alpha$ intensity, [N$\,$II]/H$\alpha$, [S$\,$II]/H$\alpha$, and
[Ne$\,$III]/[Ne$\,$II] for our standard model with temperature 40kK.  Vertical
intensity cuts that show line ratios at $x$ offsets of 0~kpc, 0.3~kpc,
0.6~kpc, and 0.9~kpc are shown in Figure 7.  Figure 6 shows that the
[N$\,$II]/H$\alpha$ and [S$\,$II]/H$\alpha$ ratios increase with distance from
the ionizing source.  This is due to the progressive hardening of the
radiation field with increasing distance from the ionizing source
\citep[WM;][]{2003ApJ...586..902H}.  [Ne$\,$III]/[Ne$\,$II] decreases with
distance from the ionizing source due to the small number of high energy
photons available at large distances to ionize Ne$^+$ to Ne$^{2+}$ (only for
the cut at $x=0.9$ kpc do we see a hint of a rise in [Ne$\,$III]/[Ne$\,$II]
with $|z|$, but it is clear from Figure 6 that this due to the idealized
cone-like geometry of the simulations and cannot be considered a plausible
match to the data; for the same reason, spikes and dips appear in most other
line ratios at $|z|<500$ pc for $x=0.9$ kpc, in clear disagreement with the
data). [O$\,$III]/H$\alpha$ falls with $|z|$, He$\,$I/H$\alpha$ is
overpredicted and [O$\,$I]/H$\alpha$ is badly underpredicted for all cuts.
The 50 kK model does produce [O$\,$III]/H$\alpha$ rising with $|z|$, but
[Ne$\,$III]/[Ne$\,$II] ranges from 6 to 20 and falls with $|z|$, while the
situation for He$\,$I/H$\alpha$ and [O$\,$I]/H$\alpha$ is not improved.

\clearpage
\begin{figure}
\epsscale{.80}
\includegraphics[scale=.6]{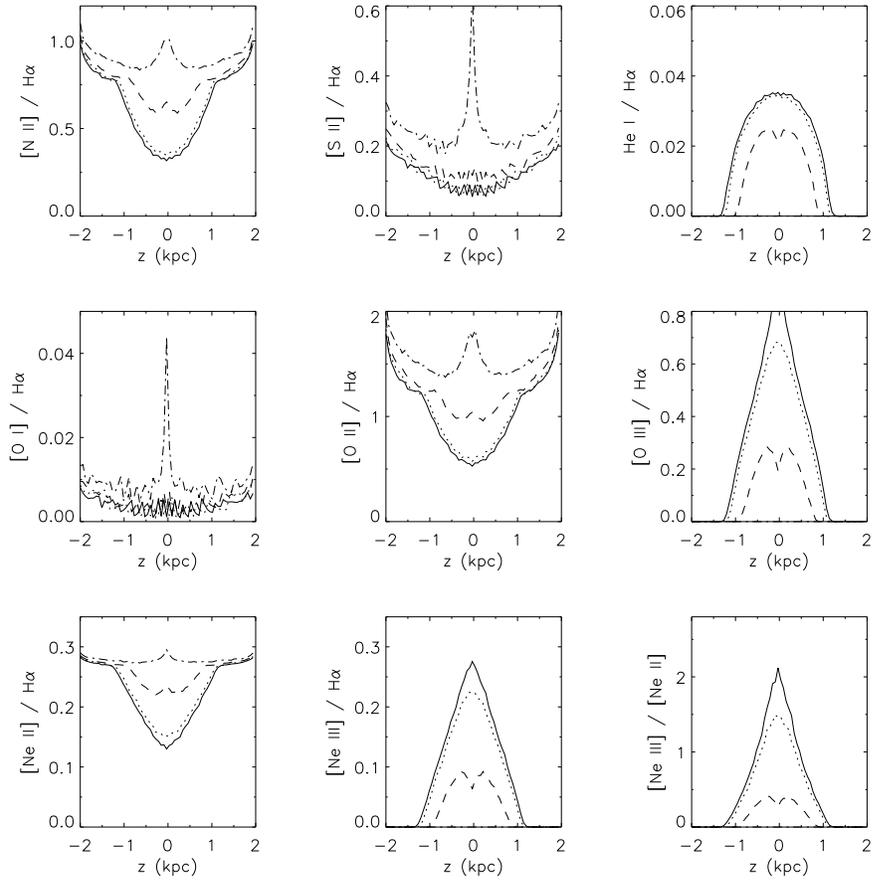}
\caption{Vertical cuts at $x=$ 0 kpc (solid), 0.3 kpc (dotted), 0.6 kpc
  (dashed) and 0.9 kpc (dot-dashed) showing the variation of line ratios with
  $z$ for the 40 kK leaky model with luminosity $10^{51}$ H-ionizing photons
  per second.
\label{fig8}}
\end{figure}
\clearpage

The leaky 40kK model (Figure 8) shows falling [Ne$\,$III]/[Ne$\,$II] with
$|z|$ for all cuts.  At $x=0.9$, the values are at least comparable to the
data.  [O$\,$III]/H$\alpha$ and He$\,$I/H$\alpha$ still fall with $|z|$,
again disagreeing with the data.  Boosting the ionizing luminosity badly
overpredicts [Ne$\,$III]/[Ne$\,$II] and this ratio falls with $|z|$ between 0
and 1 kpc, while all optical ratios are badly matched.  The leaky,
high-luminosity 40kK model best matches the run of He$\,$I/H$\alpha$ (see
discussion in WM), but again shows a neon ratio falling with $|z|$ and fails
badly on the other optical ratios.  In all the models considered here,
[S$\,$III]/[Ne$\,$II] (not shown) is badly overpredicted.

In summary, while the models can semi-quantitatively reproduce the behavior of
[N$\,$II]/H$\alpha$ and [S$\,$II]/H$\alpha$, they generally overpredict
[Ne$\,$III]/[Ne$\,$II] and do not predict that this ratio should rise with
$|z|$.  Only at large distances in $x$ from the ionizing source do the values
begin to approach those observed (yet they still fall with $|z|$ in the
model).  The models also do not predict a rising [O$\,$III]/H$\alpha$ with
$|z|$ as observed.  The modeled [O$\,$I]/H$\alpha$ values are generally too
low, and He$\,$I/H$\alpha$ is not well matched.  These results demonstrate
that hardening of the radiation field is insufficient to explain the data.  An
extra source of non-ionizing heating will also not affect the neon ratio as it
is insensitive to gas temperature.

One potential explanation for enhanced [Ne$\,$III]/[Ne$\,$II] in the halo
would be the creation of higher levels of ionization due to increased
collisional ionization.  Such an enhancement could arise behind shocks in the
halo.  Although such shocks could not dominate the overall energetics of the
diffuse ionized layers because of the large power requirements, they could
provide the extra ionization at large heights above the disk.  However, there
are several difficulties with comparing our results to shock models.  First,
published models of steady-state shocks do not currently including predictions
for infrared emission line intensities. Second, since these shocks are taking
place in a low density medium, the length scales of the postshock cooling and
recombination zones can be shown to approach the scale-height of the DIG. In
such a case, it is necessary to consider the possibility of "truncated shocks"
\citep{1993ARA&A..31..373D} in which the integrated quantities are affected by
the endpoint chosen for the recombination zone.  Finally, given the hard
radiation field produced by the postshock gas, the majority of the [Ne$\,$II]
and [Ne$\,$III] that forms occurs in the postshock recombination zone.  Thus
the column densities should be sensitive to the presence and spectrum of the
external radiation field.  A more complete investigation of the sensitivity
and role of shocks in affecting the ionization structure at high latitude will
be included in subsequent investigations.

\section{Conclusions}

We have presented {\it Spitzer} IRS spectra of the disk and halo ($z=\pm 1$
kpc) of the edge-on spiral galaxy NGC 891.  Our most important result is that
the [Ne$\,$III]/[Ne$\,$II] ratio is found to be enhanced in the halo relative
to the disk, and this is very difficult to reproduce in pure photo-ionization
models.

Even this single observation presents a significant problem for models of DIG
ionization featuring massive stars as the sole ionizing source.  In a future
paper we will explore shocks as a second source of ionization, but we remind
the reader that other processes, such as cooling hot gas or turbulent mixing
layers, may well be relevant and deserve further study.

We have detected several PAH features in the disk and the halo.  If in
exponential layers, their scale-heights are in the rough range $330-530$ pc,
probably between the CO and HI scale-heights.  However, extinction in the disk
may reduce these scale-heights significantly.  Most have similar halo-disk
emission contrasts, suggesting little variation in the PAH population, as
expected for low starlight intensities.  There is also very little halo-disk
contrast in EW for most of the features, putting limits on modification
between disk and halo of the relative numbers of PAHs and the very small
grains responsible for the continuum.  Only the 17.4 $\mu$m feature stands out
as having a higher halo-disk contrast and a higher EW than the others,
possibly indicating a preference for larger PAHs in halos, although the origin
of this feature is uncertain.  The information on PAHs in halos is limited so
far, but such work on the vertical variations of the various emission features
has great potential for understanding how interstellar processes can transport
grains and modify their populations.

Our understanding of both the ionized gas and dust phases in halos will
benefit from additional spectroscopic data in a range of halo environments,
and we are in the process of acquiring such data with the IRS (program ID
40284; PI: R. Rand).  In a future paper we will extend our NGC 891 data set
with spectra at $z=\pm 2$ kpc, examine the halo of the more actively
star-forming NGC 5775, where the rise of [O$\,$III]/H$\alpha$ with $|z|$ is
even more problematic for pure photoionization models, and also the halo of
NGC 3044, where a secondary ionization source is not evidently required, as
[O$\,$III]/H$\alpha$ clearly falls with $|z|$ \citep{2000A&A...362..119T}.

%% Included in this acknowledgments section are examples of the
%% AASTeX hypertext markup commands. Use \url without the optional [HREF]
%% argument when you want to print the url directly in the text. Otherwise,
%% use either \url or \anchor, with the HREF as the first argument and the
%% text to be printed in the second.

\acknowledgments

We thank G. Stacey for a brief but useful discussion about extinction in the
mid-IR.  We thank an anonymous referee for extensive comments which helped the
paper significantly.  This work is based (in part) on observations made with
the Spitzer Space Telescope, which is operated by the Jet Propulsion
Laboratory, California Institute of Technology under a contract with
NASA. Support for this work was provided by NASA through an award issued by
JPL/Caltech.

%% To help institutions obtain information on the effectiveness of their
%% telescopes, the AAS Journals has created a group of keywords for telescope
%% facilities. A common set of keywords will make these types of searches
%% significantly easier and more accurate. In addition, they will also be
%% useful in linking papers together which utilize the same telescopes
%% within the framework of the National Virtual Observatory.
%% See the AASTeX Web site at http://www.journals.uchicago.edu/AAS/AASTeX
%% for information on obtaining the facility keywords.

%% After the acknowledgments section, use the following syntax and the
%% \facility{} macro to list the keywords of facilities used in the research
%% for the paper.  Each keyword will be checked against the master list during
%% copy editing.  Individual instruments can be provided in parentheses,
%% after the keyword, but they will not be verified.

Facilities: \facility{Spitzer Space Telescope}.

%% The reference list follows the main body and any appendices.
%% Use LaTeX's thebibliography environment to mark up your reference list.
%% Note \begin{thebibliography} is followed by an empty set of
%% curly braces.  If you forget this, LaTeX will generate the error
%% "Perhaps a missing \item?".
%%
%% thebibliography produces citations in the text using \bibitem-\cite
%% cross-referencing. Each reference is preceded by a
%% \bibitem command that defines in curly braces the KEY that corresponds
%% to the KEY in the \cite commands (see the first section above).
%% Make sure that you provide a unique KEY for every \bibitem or else the
%% paper will not LaTeX. The square brackets should contain
%% the citation text that LaTeX will insert in
%% place of the \cite commands.

%% We have used macros to produce journal name abbreviations.
%% AASTeX provides a number of these for the more frequently-cited journals.
%% See the Author Guide for a list of them.

%% Note that the style of the \bibitem labels (in []) is slightly
%% different from previous examples.  The natbib system solves a host
%% of citation expression problems, but it is necessary to clearly
%% delimit the year from the author name used in the citation.
%% See the natbib documentation for more details and options.

%%\begin{thebibliography}{}
\bibliography{ms}
%%\end{thebibliography}

\clearpage

%% Tables should be submitted one per page, so put a \clearpage before
%% each one.

%% Two options are available to the author for producing tables:  the
%% deluxetable environment provided by the AASTeX package or the LaTeX
%% table environment.  Use of deluxetable is preferred.
%%

%% Three table samples follow, two marked up in the deluxetable environment,
%% one marked up as a LaTeX table.

\begin{deluxetable}{llllll}
\tabletypesize{\scriptsize}
\tablecaption{NGC 891 IRS Observations\tablenotemark{a}\label{tbl-1}}
\tablewidth{0pt}
\tablehead{
\colhead{AOR} &
\colhead{Date (2006)}   & 
\colhead{Pointing\tablenotemark{b}} & 
\colhead{R.A. (J2000.0)} & 
\colhead{Decl. (J2000.0)} & 
\colhead{Integ. time per nod (sec)}
}
\startdata
14595584 & March 4 & Disk      & 2$^{\rm h}$ 22$^{\rm m}$ 36.8$^{\rm s}$ & 42$^{\circ}$ $22\arcmin$ 27$\arcsec$ & 2x480 \\
14595840 & March 4 & Sky       & 2$^{\rm h}$ 23$^{\rm m}$  6.4$^{\rm s}$ & 42$^{\circ}$ $22\arcmin$ 27$\arcsec$ & 1x480 \\
14596096 & March 4 & Halo East & 2$^{\rm h}$ 22$^{\rm m}$ 38.6$^{\rm s}$ & 42$^{\circ}$ $22\arcmin$ 22$\arcsec$ & 9x480 \\
14596352 & March 4 & Halo East & 2$^{\rm h}$ 22$^{\rm m}$ 38.6$^{\rm s}$ & 42$^{\circ}$ $22\arcmin$ 22$\arcsec$ & 9x480 \\
14596608 & Feb 2\tablenotemark{C}   & Halo East & 2$^{\rm h}$ 22$^{\rm m}$ 38.6$^{\rm s}$ & 42$^{\circ}$ $22\arcmin$ 22$\arcsec$ & 7x480 \\
14596864 & March 4 & Halo West & 2$^{\rm h}$ 22$^{\rm m}$ 35.0$^{\rm s}$ & 42$^{\circ}$ $22\arcmin$ 38$\arcsec$ & 9x480 \\
14597120 & March 4 & Halo West & 2$^{\rm h}$ 22$^{\rm m}$ 35.0$^{\rm s}$ & 42$^{\circ}$ $22\arcmin$ 38$\arcsec$ & 9x480 \\
14597376 & March 4 & Halo West & 2$^{\rm h}$ 22$^{\rm m}$ 35.0$^{\rm s}$ & 42$^{\circ}$ $22\arcmin$ 38$\arcsec$ & 7x480 \\

\enddata

%% Text for table notes should follow after the \enddata but before
%% the \end{deluxetable}. Make sure there is at least one \tablenotemark
%% in the table for each \tablenotetext.

\tablenotetext{a}{Spitzer program ID 20380}
\tablenotetext{b}{Pointing center of the field of view (each nod
pointing is symmetrically offset from this position)}
\tablenotetext{C}{These data are not used}

\end{deluxetable}
\clearpage

\begin{deluxetable}{lcccccc}
\tabletypesize{\scriptsize}
\tablecaption{Infrared Line Intensities and PAH Equivalent Widths\label{tbl-2}}
\tablewidth{0pt}
\tablehead{
\colhead{} & \colhead{Disk} & \colhead{Disk} & \colhead{Halo East} & \colhead{Halo East} & \colhead{Halo West} &\colhead{Halo West} \\ 
\colhead{Line} & \colhead{Intensity\tablenotemark{a}} & \colhead{EW} & \colhead{Intensity} & \colhead{EW} & \colhead{Intensity} & \colhead{EW} \\
\colhead{} &
\colhead{} &
\colhead{($\mu$m)} &
\colhead{} &
\colhead{($\mu$m)} &
\colhead{} &
\colhead{($\mu$m)} \\
}
\startdata
\phd [S$\,$IV] 10.51 $\mu$m   & $<$1.9\tablenotemark{b} &   & $<$1.0  &  & $<$1.1 &  \\
\phd [Ne$\,$II] 12.81 $\mu$m  & $320 \pm 20$ &  & $12 \pm 1$    &  & $11 \pm 1 $    &  \\
\phd [Ne$\,$III] 15.56 $\mu$m & $41 \pm 3$   &  & $3.8 \pm 0.4$ &  & $3.4 \pm 0.3$ &  \\
\phd [S$\,$III] 18.71 $\mu$m  & $87 \pm 6$   &  & $3.0 \pm 0.3$ &  & $3.0 \pm 0.3$ &  \\
\phd H$_2$ {\it S}(1) $J=3-1$ 17.03 $\mu$m  & $49 \pm 3$  &  & $4.0 \pm 0.3$ &  & $4.3 \pm 0.3$ & \\
\phd PAH\tablenotemark{c} 11.2 $\mu$m          & $1400 \pm 100 $ & $0.85 \pm 0.04$ & $105 \pm 9$ & $0.99 \pm 0.05$ & $ 89 \pm 7 $ & $0.77 \pm 0.04$ \\ 
\phd PAH 12.0 $\mu$m          & $85 \pm 7 $ & $0.043 \pm 0.002$ & $2.5 \pm 0.9$ & $0.02 \pm 0.01$ & $ 4.7 \pm 0.8 $ & $0.04 \pm 0.01$ \\ 
\phd PAH 12.7 $\mu$m          & $600 \pm 50 $ & $0.32 \pm 0.02$ & $36 \pm 3$ & $0.36 \pm 0.02$ & $ 23 \pm 2 $ & $0.21 \pm 0.01$ \\ 
\phd PAH 16.5 $\mu$m          & $68 \pm 5 $ & $0.08 \pm 0.01$ & $4.0 \pm 0.5$ & $0.08 \pm 0.01$  & $ 3.6 \pm 0.4 $ & $0.08 \pm 0.01$ \\ 
\phd PAH 17.4 $\mu$m          & $32 \pm 2 $ & $0.038 \pm 0.002$ & $5.5 \pm 0.6$ & $0.12 \pm 0.01$ & $ 3.6 \pm 0.4 $ & $0.08 \pm 0.01$ \\

\enddata

%% Text for table notes should follow after the \enddata but before
%% the \end{deluxetable}. Make sure there is at least one \tablenotemark
%% in the table for each \tablenotetext.

\tablenotetext{a}{Units are (10$^{-17}$ erg cm$^{-2}$ s$^{-1}$ arcsec$^{-2}$)}
\tablenotetext{b}{Upper limits are $3\sigma$.}
\tablenotetext{c}{Intensities for the 11.2 $\mu$m feature include the secondary peak at about 11.0 $\mu$m.}

\end{deluxetable}

\begin{deluxetable}{lccccc}
\tabletypesize{\scriptsize}
\tablecaption{Infrared Line Intensities and PAH Equivalent Widths for GTO Data\tablenotemark{a}\label{tbl-3}}
\tablewidth{0pt}
\tablehead{
\colhead{} & 
\colhead{A} & \colhead{B} & \colhead{C} & \colhead{D} & \colhead{E} \\
}
\startdata
\phd & & & & & \\
\phd R.A. (J2000.0) & 2$^{\rm h}$ 22$^{\rm m}$ 34.7$^{\rm s}$ & 2$^{\rm h}$ 22$^{\rm m}$ 35.5$^{\rm s}$& 2$^{\rm h}$ 22$^{\rm m}$ 36.5$^{\rm s}$ & 2$^{\rm h}$ 22$^{\rm m}$ 37.6$^{\rm s}$ & 2$^{\rm h}$ 22$^{\rm m}$ 38.4$^{\rm s}$\\
\phd Decl. (J2000.0) & 42$^{\circ}$ $20\arcmin$ $57\arcsec$ & 42$^{\circ}$ $21\arcmin$ $50\arcsec$ & 42$^{\circ}$ $22\arcmin$ $19\arcsec$  & 42$^{\circ}$ $22\arcmin$ $46\arcsec$  & 42$^{\circ}$ $23\arcmin$ $14\arcsec$ \\
\phd & & & & & \\
\phd Gas phase intensities\tablenotemark{b} & & & & & \\
\phd & & & & & \\
\phd [S$\,$IV]   10.51 $\mu$m  & $<$25\tablenotemark{c} & $<$25 & $<$25 &  $<$25 & $<$25 \\
\phd [Ne$\,$II]  12.81 $\mu$m  & $170 \pm 14$ & $170 \pm 14$ & $230 \pm 18$ & $510 \pm 40 $ & $90 \pm 7$ \\
\phd [Ne$\,$III] 15.56 $\mu$m  & $21 \pm 4$ &  $23 \pm 4$ & $32 \pm 4$ & $66 \pm 5$ & $<$12  \\
\phd [S$\,$III]  18.71 $\mu$m  & $40 \pm 4$ &  $43 \pm 4$ & $64 \pm 5.3$ & $180 \pm 30$ & $30 \pm 4$   \\
\phd [S$\,$III]  33.48 $\mu$m  & $93 \pm 6$ &  $150 \pm 10$ & $170 \pm 12$ & $280 \pm 19$ & $81 \pm 6$  \\
\phd [Si$\,$II]  34.72 $\mu$m  & $160 \pm 12$  &  $220 \pm 15$ & $260 \pm 20$ & $330 \pm 20$ & $100 \pm 7$  \\
\phd H$_2$ {\it S}(1) $J=3-1$ 17.03 $\mu$m & $41 \pm 3$  &  $36 \pm 3$ & $40 \pm 3$ & $51 \pm 4$ & $24 \pm 2$ \\
\phd & & & & & \\
\phd PAHs intensities\tablenotemark{b} and EWs ($\mu$m)\\
\phd & & & & & \\
\phd PAH\tablenotemark{d} 11.2 $\mu$m  & $940 \pm 80$  & $750 \pm 60$ & $ 1000 \pm 80$  & $1300 \pm 100$ & $740 \pm 60$ \\ 
\phd & $1.04 \pm 0.05$ & $0.52 \pm 0.03$ & $0.65 \pm 0.03$ & $0.58 \pm 0.03$  & $0.95 \pm 0.05$ \\
\phd PAH 12.0 $\mu$m & $<42$  & $<42$  & $<42$  & $<42$  & $<42$  \\
\phd & $<0.05$ & $<0.03$ & $<0.03$ & $<0.03$ & $<0.05$ \\
\phd PAH 12.7 $\mu$m & $380 \pm 30$ & $450 \pm 40$  & $ 570 \pm 45$  & $700 \pm 60$  & $210 \pm 17$  \\
\phd &  $0.29 \pm 0.01$ & $0.31 \pm 0.02$ & $0.39 \pm 0.02$ & $0.36 \pm 0.02$ & $0.245 \pm 0.01$ \\
\phd PAH 16.5 $\mu$m & $<19$  & $42 \pm 8$  & 
$ 53 \pm 8$  & $42 \pm 8$  & $<19$ \\ 
\phd & $<0.05$ & $0.07 \pm 0.01$ & $0.09 \pm 0.01$ & $0.04 \pm 0.01$  & $<0.05$\\
\phd PAH 17.4 $\mu$m & $<17$  & $<17$ & $<17$  & $<17$  & $<17$  \\ 
\phd & $<0.05$ & $<0.03$ & $<0.03$& $<0.02$& $<0.06$\\

\enddata

%% Text for table notes should follow after the \enddata but before
%% the \end{deluxetable}. Make sure there is at least one \tablenotemark
%% in the table for each \tablenotetext.

\tablenotetext{a}{Spitzer program ID 97}
\tablenotetext{b}{Units are (10$^{-17}$ erg cm$^{-2}$ s$^{-1}$ arcsec$^{-2}$)}
\tablenotetext{c}{Upper limits are $3\sigma$.}
\tablenotetext{d}{Intensities for the 11.2 $\mu$m feature include the secondary peak at about 11.0 $\mu$m.}

\end{deluxetable}

%% The following command ends your manuscript. LaTeX will ignore any text
%% that appears after it.

\end{document}